\documentclass[twocolumn,prl,aps,superscriptaddress,showpacs]{revtex4}
\usepackage{graphicx}
\begin{document}
\title {Generating multi-scaling networks with different types of nodes}
\author{Shi-Jie Yang}
\author{Hu Zhao}
\affiliation{Department of Physics, Beijing Normal University,
Beijing 100875, China}

\begin{abstract}
A variety of scale-free networks have been created since the
pioneer work by A.-L. Barab\'{a}si and R. Albert. All this
networks are homogeneous since they are composed of the same kind
of nodes. In the realistic world, however, one element (node or
vertex) in the network may play different roles and hence has
different functions. In this Letter, we develop a new kind of
network to account for this property. In our model, each type of
nodes may exhibit a scaling law in the degree distribution and the
scaling exponents are adjustable. As a consequence, the whole
network lacks of such scaling characteristics, which indicates
that many previous statistical results based on empirical data
that claimed to be scale-free networks may need to be reexamined.
This model poses an alternative way of the network division other
than the module method. Besides, one can expect that this new
network will exhibit some interesting properties concerning the
dynamical processes on it.

\end{abstract}
\maketitle

In recent years, an increasing interest on complex networks has
risen in physical society because it shows as one of the most
promising tools to describe various social, biological, and
technical systems, such as the
Internet\cite{Caldarelli,Pastor1,Pastor2}, the World Wide Web
(WWW)\cite{Albert} or collaboration networks\cite{Newman,Jeong}.
In the models, vertices or nodes may represent people, proteins,
species, routers, or html documents while the links between the
nodes correspond to acquaintance, physical interactions, predation
relationships, cable connections, or hyper-links, respectively. In
real-world networks, two outstanding phenomenon are mostly
presented. One is the small-world
phenomenon\cite{Watts,Jespersen}, which refers to that one can
promptly reach remote parts of the network through very few hops.
The other is the scale-free phenomenon\cite{BA,AB,Garl}, which
refers to the degree of the nodes (the number of links connected
to it) obeying a power-law distribution as $P(k)\sim k^{-\gamma}$.

To explain the scale-free property of a network, A.-L.
Barab\'{a}si and R. Albert (BA)\cite{BA} designed a model by
introducing the concepts of growing network and of preferential
attachment. The network are growing at a constant rate and new
nodes are attached to the older ones with a probability which is
proportional to the degrees of the preexisting nodes,
\begin{equation}
\Pi (k_i)=\frac{k_i}{\sum_j (k_j)}.
\end{equation}
The network generated in this way has a fixed scaling exponent
$\gamma=3$. Variant models that assign a fitness to every
newly-added node, which accounts for the importance of the node
for attracting new links, can generate networks with adjustable
scaling exponents\cite{Cald,Bog,Barrat}. Besides, many other
network models and methods are proposed to probe the various
properties in real-world
networks\cite{Doro,Ravasz1,Ravasz2,Girvan,Newman2}.

Nevertheless, all the previous works are focused on networks in
which all the nodes are identical and play a unique function in
the homogeneous (or inhomogeneous) network. As one explicitly
examines into a real-world network, however, it can be found that
a node is generally not play only one role in the network. It may
play multi-roles or couple with nodes in other systems. As show in
Fig.1, nodes of type A or type B respectively form two subsystems
with links between some of them. From another point of view, if we
regard the whole object as two different networks which are
respectively consisted of type A or type B nodes, then a node in
one network may couple with a node in the other network even
though this two networks may have completely different topology or
function. These nodes may behavior quite differently as they
present in different environments. For example, a researcher may
work in two or more fields and cooperate with different authors.
He writes papers in condensed matter physics with collaborators
who are always engaged in this field while he may also contribute
to the high energy physics by participating another group. He may
also have an independent friendship network\cite{Palla}.

Many works have been carried out to deal with the community or
module structure of the network, where nodes is liable to
interconnect within one
community\cite{Ravasz1,Girvan,Palla,Clauset}. In this Letter, we
introduce another way to catalogue the network nodes according to
their functions as relative independent subsystems. We will
construct a heterogeneous network consisted of two types of nodes,
type A and type B. As in a biochemical reaction processes, the
reactants and their relationship form a complex network. One type
of reactant preferentially combines with some other specific
reactants. The attractance or binding energy $\alpha_1$ between
the same type of nodes (A-A or B-B) differs from the attractance
$\alpha_2$ between the different types of nodes (A-B). When a new
node (type A or type B) joins the growing network, two rules
govern its attachment to the preexisting nodes: i) The
Barab\'{a}si and Albert preferential attachment, and ii) The
selection of partners according to its genus. The whole network is
such composed of two types of nodes or vertices, {\it i.e.}, it is
a heterogeneous network with two connected subgraphs of $G_A$ and
$G_B$. Figure 1 is a sample network generated by this algorithm
for 30 nodes with $\alpha_2/\alpha_1=0.1$ and the proportion of
type A nodes $p_A=0.6$.

It is found that our model network exhibits a multi-scaling
structure in which the degree distribution for each subgraph $G_A$
or $G_B$ shows a scaling law. As a consequence, the total degree
distribution lacks of such scaling characteristics except for some
special cases. The result is instructive for many recent
statistics on empirical data that claimed to have the scaling law
distribution. They are generally not reliable if the nodes play
multi-roles in the generating process or the network is composed
of several subsystems and need to be carefully reexamined. Many of
them may show as the pseudo-scaling laws.

We assume that the newly-added node presents as type A with
probability $p_A$ and as type B with probability $p_B=1-p_A$,
respectively. Each node has $m$ feet to be connected to the
existing network. The vertices of type A and type B respectively
form two vertex sets of $V_A=\{v_{1},v_{2},\cdots,v_{N_A}\}$ and
$W_B=\{w_{1},w_{2},\cdots,w_{N_B}\}$. Here $N_A$ and $N_B$ are the
total numbers of node A and node B, respectively. From the above
rules, we obtain the continuously growing equations
\begin{widetext}
\begin{eqnarray}
\frac{\partial k_A(i)}{\partial t}&=&\frac{m p_A \alpha_1
k_A(i)}{\sum_{j\in V_A} \alpha_1 k_A(j)+\sum_{j\in W_B} \alpha_2
k_B(j)}+\frac{m p_B \alpha_2 k_A(i)}{\sum_{j\in V_A} \alpha_2
k_A(j)+\sum_{j\in W_B} \alpha_1 k_B(j)}  \nonumber\\
\frac{\partial k_B(i)}{\partial t}&=&\frac{m p_A \alpha_2
k_B(i)}{\sum_{j\in V_A} \alpha_1 k_A(j)+\sum_{j\in W_B} \alpha_2
k_B(j)}+\frac{m p_B \alpha_1 k_B(i)}{\sum_{j\in V_A} \alpha_2
k_A(j)+\sum_{j\in W_B} \alpha_1 k_B(j)}, \label{grow}
\end{eqnarray}
\end{widetext}
where $k_A(i)$ and $k_B(i)$ are the degree of the $i$-th node of
either type A or type B, respectively. The first line of the above
equations describes the growth rate of type A nodes in the
network. The first term on the right-handed side represents a
newly-added type A node, which is generated with probability
$p_A$, is attached to a type A node in the preexisting network
while the second term is a newly-added type B node to be attached
also to a type A node in the network. Analogously, the second line
describes the growth rate of type B nodes in the network.

To solve the above equations, we consider the thermodynamical
approximation. Suppose the system has multi-scaling law, namely,
the dynamic exponents depend on the attractance $\alpha_1$ and
$\alpha_2$,
\begin{eqnarray}
k_A(t,t_0,i)&=&m(\frac{t}{t_0})^{\beta_A}\nonumber\\
k_B(t,t_0,i)&=&m(\frac{t}{t_0})^{\beta_B},
\end{eqnarray}
where $t_0$ is the time at which the node $i$ was born. The
dynamic exponents $\beta_A$ and $\beta_B$ are bounded, i.e.
$0<\beta_A,\beta_B<1$ because a node always increases the number
of links in time ($\beta_A,\beta_B>0$) and $k_A,k_B$ cannot
increase faster than $t$ ($\beta_A,\beta_B<1$). We calculate the
sum over $j$ in Eqs.(\ref{grow}) by writing them in the integral
forms. By noting that the nodes respectively belong to sets $V_A$
and $W_B$ with probability $p_A$ and $p_B$, one has
\begin{eqnarray}
\sum_{j\in V_A}k_A(j)&=& m\int_1^t p_A(\frac{t}{t_0})^{\beta_A}
\stackrel{t\rightarrow\infty}{=}\frac{mp_A
t}{1-\beta_A}\nonumber\\
\sum_{j\in W_B} k_B(j)&=& m\int_1^t p_B(\frac{t}{t_0})^{\beta_A}
\stackrel{t\rightarrow\infty}{=}\frac{mp_A t}{1-\beta_B}.
\label{kab}
\end{eqnarray}
Here we have dropped the $t^\beta$ term for it becomes less
important as $t\rightarrow\infty$.

After some calculations by introducing a variable
$z=(1-\beta_A)/(1-\beta_B)$, we obtain a third order equation as
\begin{widetext}
\begin{equation}
z^3+[\frac{p_A}{p_B}(\frac{2\alpha_1}{\alpha_2}+\frac{\alpha_2}{\alpha_1}-1)
+(\frac{\alpha_2}{\alpha_1}-2)]z^2
+[\frac{p_A}{p_B}(-\frac{2\alpha1}{\alpha_2}-\frac{\alpha_2}{\alpha_1}+1)
+\frac{p_A^2}{p_B^2}(2-\frac{\alpha_2}{\alpha_1})]z-\frac{p_A^2}{p_B^2}=0,
\end{equation}
\end{widetext}
and
\begin{equation}
1-\frac{1}{\frac{p_A\alpha_1}{p_A\alpha_1+p_B\alpha_2
z}+\frac{p_B\alpha_2}{p_B\alpha_1 z+p_A\alpha_2}+1}=\beta_A.
\end{equation}
Therefore, it demonstrates that in the thermodynamical limit, the
degree distributions of the constituent nodes respectively obey
the power law. The power exponents of each type of nodes are
generally different. Obviously, the total degree distribution that
comprises both type A and type B nodes will not exhibit the
scaling characteristics.

Figure 2 plots the dependence of the exponents $\gamma_A$ and
$\gamma_B$ ($\gamma=1+1/\beta$) on the relative attractive
strength between the nodes $\alpha_2/\alpha_1$. It is seen that
the exponents can be either larger than 3 or in the regime of
$2<\gamma<3$, depending on the parameter value of
$\alpha_2/\alpha_1$. The upper panel shows that if $\gamma_A>3$,
then $\gamma_B<3$ whereas if $\gamma_A<3$, then $\gamma_B>3$. For
smaller $\alpha_2/\alpha_1$, the newly-added node will
preferentially attaches to the same type of preexisting nodes. In
the case of $p_A>0.5$, type A nodes with larger degrees are
continuing to attract more links and thus the exponent is relative
small. As $\alpha_2/\alpha_1\rightarrow 0$, the scaling exponents
for both genuses of nodes tend to 3. In this case, the coupling
between the two types of nodes are weak and the resultant network
nearly divides into two isolated subsystems. At the point of
$\alpha_2/\alpha_1=1$, which corresponds to the case that A and B
nodes are completely identical, our model degenerates into the
ordinary BA model with a fixed exponent $\gamma=3$. As a special
case of $p_A=p_B=0.5$, the exponents $\gamma_A=\gamma_B\equiv 3$,
regardless of the detailed values of $\alpha_2/\alpha_1$.

For larger $\alpha_2/\alpha_1$ ($\alpha_2/\alpha_1>1$, see the
lower panel of Fig.2) where one type of node will preferentially
attach to the different type of nodes (A to B or B to A), the
network seems to be interweaved alternatively by node A and node
B. Hence for $p_A>0.5$, it becomes hard for the type A nodes that
already have more links to get even more links, i.e., the degree
growth is damped. Most nodes of this type will remain few links
and so the scaling exponent becomes larger. On the other hand, for
$p_B<0.5$, those nodes of type B that already have more links will
continue to attract more links from type A nodes and so the
scaling exponent decreases as $\alpha_2/\alpha_1$ increases.

To check the above theoretical results, we simulate the growth
process on the computer. We start with $N_0=4$ interconnected
nodes consisted of 2 type A and 2 type B nodes, respectively.
Figure 3 displays the simulated degree distribution $P(k)$ for
$p_A=0.9$ and $\alpha_2/\alpha_1=0.5$. A total of 200,000 nodes
are involved. The solid lines are respectively the theoretical
results for type A ($\gamma_A=2.95$) and type B ($\gamma_B=4.06$)
nodes. It is seen that the simulating data coincide with the
theoretical result quite well. Here we point out an important
fact. Just from the simulating data (see the inset of Fig.3), one
may misinterpret that the total distribution indiscriminating the
node genuses also show a power law relation. However, it is
incorrect from the theoretical considerations. This fact
precautions us that it should be wary when dealing with empirical
data that include several types of nodes to draw out a scaling
law.

Figure 4 is the same as in Fig.3 for $\alpha_2/\alpha_1=3.0$. The
two solid lines correspond to the theoretical results with
$\gamma_A=3.96$ and $\gamma_B=2.18$, respectively. It shows that
for type B nodes, the data deviate form the theoretical lines.
This is because we have dropped in Eqs.(\ref{kab}) a term which
behaviors as $t^{\beta}$. When $\beta\rightarrow 1$ (or
$\gamma\rightarrow 2$), as in the present case, this term becomes
increasingly important for limited number of simulating nodes and
hence the approximation becomes poor. With the increase of node
number, the coincidence should improve correspondingly.

In summary, we have developed a bipartite network in which the
nodes are divided into two genuses according to the interactions
between them. There is an important difference between our model
and the fitness model. In the fitness model, each node has fixed
fitness while in our model, the interaction between a node with
others is dependent on the types of its partners. Just as in a
library, one classifies the books into catalogues and the books
are connected by the cross-index table. One can reach a specific
book through different ways by following the classification
method. Two major conclusions can be reached in our model: i) The
network is catalogued by the genuses or functions of the nodes
while most previous works divide the network by communities or
modules. Our model is in fact heterogenous. ii) There are
multi-scaling characteristics for each type of nodes, which
implies that the total degree lacks of a power law distribution
and many previous empirical statistics may need to be reexamined.
Our model may provide a prototype to discuss couplings between two
or even more subsystems which have scaling properties in degree
distributions. It is also interesting to explore various dynamical
processes such as searching processes\cite{Yang} on this network.
we expect that new algorithms based on our model or its possible
variants will largely promote the searching efficiency on this
kind of networks.

\centerline {Figure Captions}

Figure1 (Color online) A heterogeneous network generated from the
rules described in the context. The dotted lines represent
interactions between different types of nodes.

Figure 2 (Color online) Dependence of the scaling exponents of the
degree distributions for either type A ($\gamma_A$) or type B
($\gamma_B$) nodes on parameter ratio $\alpha_2/\alpha_1$. Upper
panel: $p_A=0.9$. Lower panel: $\gamma_A$ versus
$\alpha_2/\alpha_1$ for $p_A=0.05,0.20,0.50,0.70,0.90$.

Figure 3 (Color online) Comparison of simulating degree
distributions for type A and type B nodes with theoretical
results. $\alpha_2/\alpha_1=0.5$. The total number of nodes are
200,000 and $p_A=0.9$. Inset: The degree distribution
indiscriminating the node genus.

Figure 4 (Color online) Same as in figure 3 for
$\alpha_2/\alpha_1=3.0$.

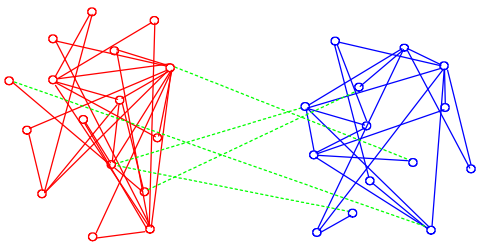
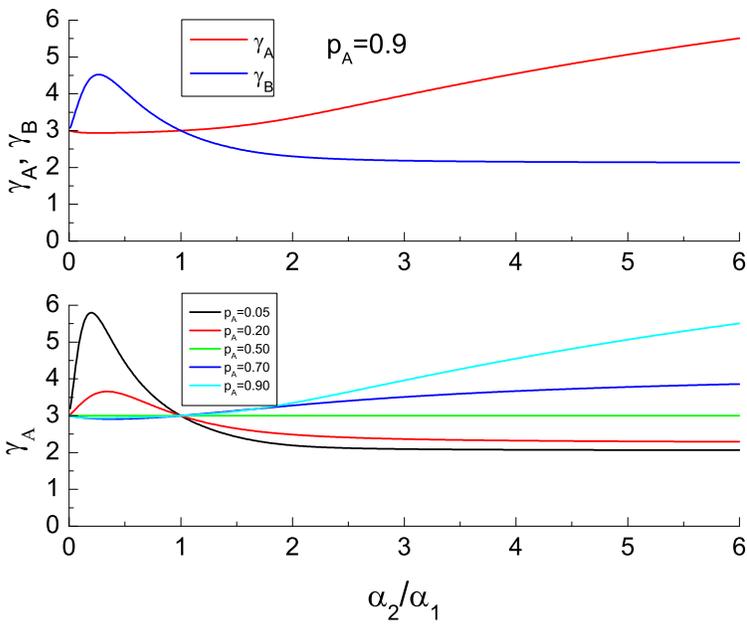
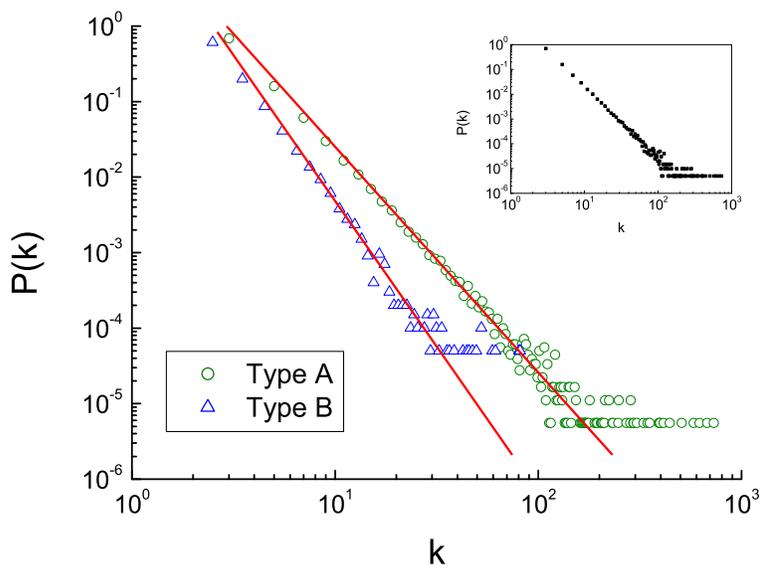
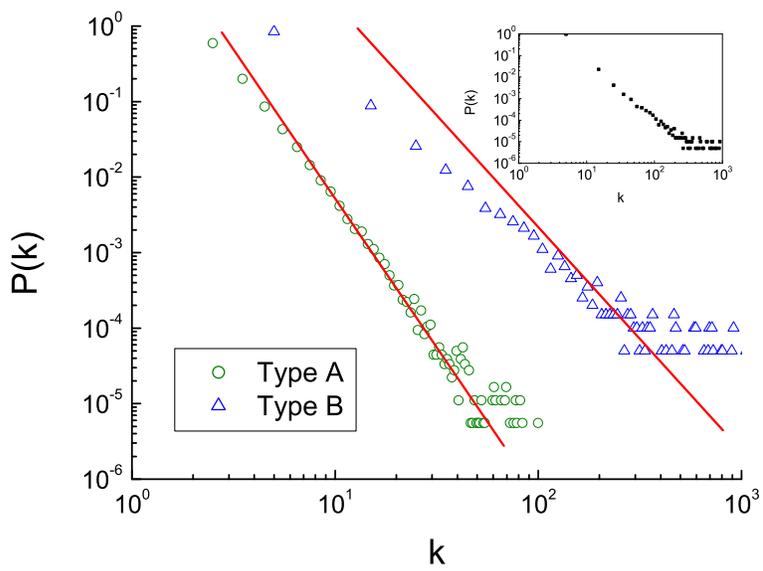
\end{document}